\documentclass[aps,preprint,showpacs]{revtex4}
\usepackage{amsfonts}
\usepackage{amsmath}
\usepackage{amssymb}
\usepackage{graphicx}
\begin{document}

\title{Fermion localization on two-field thick branes}
\author{L.B. Castro}\email[ ]{benito@feg.unesp.br}
\affiliation{Departamento de F\'{\i}sica e Qu\'{\i}mica,
Universidade Estadual Paulista, 12516-410 Guaratinguet\'{a}, S\~ao
Paulo, Brazil} 


\begin{abstract}

In a recent paper published in this journal, Almeida and collaborators [Phys. Rev. D \textbf{79}, 125022 (2009)] analyze the issue of fermion localization of fermions on a brane \,constructed from two scalar fields coupled with gravity (Bloch brane model). In that meritorious research the simplest Yukawa coupling $\eta\bar{\Psi}\phi\chi\Psi$ was considered. In that work does not analyze the zero mode in details. In this paper, the localization of fermions on two-field thick branes is reinvestigated. It is found that the simplest Yukawa coupling does not support the localization of fermions on the brane. In addition, the problem of fermion localization for some other Yukawa couplings are analyzed. It is shown that the zero mode for left-handed and right-handed fermions can be localized on the brane depending on the values for the coupling constant $\eta$ and the Bloch brane%
\'{}%
s parameter $a$.
\end{abstract}

\maketitle

\section{Introduction}

The authors of Ref. \cite{casa} analyzed the question of fermion localization on the brane \,constructed from two scalar fields coupled with gravity. That work considered the Bloch brane model \cite{ba}-\cite{du}. It was found that the Yukawa coupling $\eta\bar{\Psi}\phi\chi\Psi$, where $\eta$ is the coupling constant, allowed left-handed fermions to posses a zero mode that localizes on the brane. Fermionic resonances for both chiralities were obtained, and their appearance is related to branes with internal structure. That meritorious research did not study the zero mode in full detail. The localization problem of spin-$1/2$ fermions on thick branes is interesting and important \cite{mel}-\cite{dav}. Therefore, we believe that the conditions for obtaining normalizable zero modes on the Bloch branes model deserve to be more explored.

In Ref. \cite{mel}, a model of one-scalar brane scenario was investigated and the authors addressed the problem of chiral fermion mode confinement in thick branes. The Yukawa coupling $\eta\bar{\Psi}\phi\Psi$ was considered and the conditions for the confinement for various different Bogomol\'{}nyi-Prasad-Sommerfeld (BPS) branes, including double walls and branes interpolating between different $\mathrm{AdS}_{5}$ spacetimes were obtained. The condition stated that for appropriate values of the coupling constant, the zero mode can be normalized and one chiral fermion mode localized. In a similar context, the authors of Ref. \cite{yu} addressed the problem of fermion localization on asymmetric branes (asymmetric Bloch brane model \cite{du}). It was found that the usual couplings $\eta\bar{\Psi}(\phi+\chi)\Psi$ and $\eta\bar{\Psi}\phi\chi\Psi$ considered in the literature for two-scalar brane models do not support the localization of 4-dimensional massless fermions on the branes. Furthermore, the authors stated that the normalization of the zero mode is decided by the asymptotic behavior of the Yukawa coupling.

The main motivation of this paper is inspired on the results obtained in Ref. \cite{mel} and Ref. \cite{yu}. We reinvestigate the localization problem of fermions on two-field thick branes and we find that the usual Yukawa coupling $\eta\bar{\Psi}\phi\chi\Psi$ does not support the localization of fermions on the brane, in opposition what was adverted in Ref. \cite{casa}. In addition, we analyze the problem of fermion localization for some other Yukawa couplings and find that the zero mode for left-handed and right-handed fermions can be localized on the branes depending on the value for the coupling constant $\eta$ and the Bloch brane%
\'{}%
s parameter $a$.

\section{THE BLOCH BRANE MODEL}

The action for our system is described by \cite{wol}
\begin{equation}\label{ac1}
    S=\int d^4xdy\sqrt{|\, g|}\left[-\frac{1}{4}\,R+\frac{1}{2}\partial_{a}\phi\partial^{a}\phi+
    \frac{1}{2}\partial_{a}\chi\partial^{a}\chi-V(\phi,\chi) \right],
\end{equation}
\noindent where $g\equiv \mathrm{Det}(g_{ab})$ and the metric is
\begin{equation}\label{metric}
    ds^{2}=g_{ab}dx^{a}dx^{b}=\mathrm{e}^{2A(y)}\eta_{\mu\nu}dx^{\mu}dx^{\nu}-dy^{2},
\end{equation}
\noindent where $y=x^{4}$ is the extra dimension (the Latin indices run from $0$ to $4$), $\eta_{\mu\nu}$ is the Minkowski metric and $\mathrm{e}^{2A}$ is the so-called warp factor (the Greek indices run from $0$ to $3$). We suppose that $A=A(y)$, $\phi=\phi(y)$ and $\chi=\chi(y)$.

One can determine the static equations of motion for the above system
\begin{equation}\label{em1}
    \phi^{\prime\prime}+4A^{\prime}\phi^{\prime}=\frac{\partial V(\phi,\chi)}{\partial\phi},
\end{equation}
\begin{equation}\label{em2}
    \chi^{\prime\prime}+4A^{\prime}\chi^{\prime}=\frac{\partial V(\phi,\chi)}{\partial\chi},
\end{equation}
\begin{equation}\label{em3}
    A^{\prime\prime}=-\frac{2}{3}\left( \phi^{\prime}\,^{2}+\chi^{\prime}\,^{2} \right),
\end{equation}
\begin{equation}\label{em4}
    A^{\prime}\,^{2}=\frac{1}{6}\left( \phi^{\prime}\,^{2}+\chi^{\prime}\,^{2} \right)-\frac{1}{3}V(\phi,\chi),
\end{equation}
\noindent where prime stands for derivate with respect to $y$. We consider that the potential $V(\phi,\chi)$ can be written as \cite{wol}
\begin{equation}\label{pot1}
    V(\phi,\chi)=\frac{1}{2}\left[ \left( \frac{\partial W(\phi,\chi)}{\partial\phi} \right)^{2}+\left( \frac{\partial W(\phi,\chi)}{\partial\chi} \right)^{2} \right]
    -\frac{4}{3}\,W(\phi,\chi)^{2}
\end{equation}
\noindent where $W(\phi,\chi)$ is the superpotential and it is given by \cite{ba2}
\begin{equation}\label{sp}
    W(\phi,\chi)=2\phi-\frac{2}{3}\,\phi^{3}-2\,a\,\phi\,\chi^{2}
\end{equation}
\noindent with $a$ being a real parameter ($0<a<1/2$). The above potential leads to the following first-order differential equations, which also solve the equations of motion \cite{ba},\cite{du}, \cite{wol}
\begin{equation}\label{emp1}
   \phi^{\prime}=\frac{\partial W(\phi,\chi)}{\partial\phi},\qquad
   \chi^{\prime}=\frac{\partial W(\phi,\chi)}{\partial\chi},
\end{equation}
\begin{equation}\label{emp2}
    A^{\prime}=-\frac{2}{3}\,W(\phi,\chi),
\end{equation}

\noindent The classical solutions of the first-order differential equations (\ref{emp1}) and (\ref{emp2}) are given by
\begin{equation}\label{sol1}
    \phi(y)=\tanh(2ay),
\end{equation}
\begin{equation}\label{sol2}
    \chi(y)=\sqrt{\frac{1-2a}{a}}\,\,\mathrm{sech}(2ay),
\end{equation}
\noindent and
\begin{equation}\label{sol3}
    A(y)=\frac{1}{9a}\left\{ (1-3a)\tanh^{2}(2ay)-2\ln \left[\cosh(2ay)\right] \right\}.
\end{equation}

\begin{figure}[ht]
\begin{center}
\includegraphics[width=8cm, angle=0]{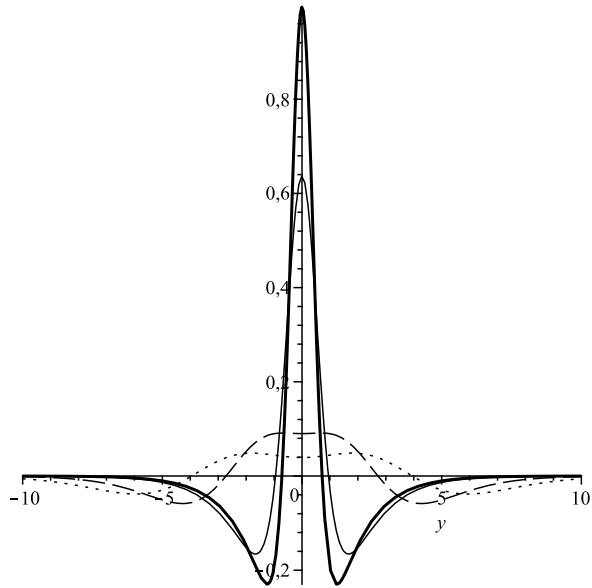}
\end{center}
\par
\vspace*{-0.1cm} \caption{The matter energy density for $a=0.09$ (dot line), $a=0.15$ (dashed line), $a=0.39$ (thin line) and $a=0.49$ (thick line).} \label{fde}
\end{figure}

The matter energy density has the form
\begin{equation}\label{de}
    \rho(y)=\mathrm{e}^{2A(y)}\left[ \frac{1}{2}\,\left( \frac{d\phi(y)}{dy} \right)^{2}+%
    \frac{1}{2}\,\left( \frac{d\chi(y)}{dy} \right)^{2}+V\left(\phi(y),\chi(y)\right) \right].
\end{equation}
\noindent The profiles of the energy density is shown in Fig. (\ref{fde}) for some values of $a$. Figure (\ref{fde}) clearly shows that the brane is localized at $y=0$ for $0.17<a<0.5$ and that for $0<a<0.17$ the behavior for the energy density shows the appearance of two peaks (two sub-branes). More detailed discussions can be found in Ref. \cite{ba}.

\section{FERMION LOCALIZATION}

The action for a Dirac spinor field coupled with the scalar fields by a general Yukawa coupling is
\begin{equation}\label{ad}
    S=\int d^{5}x\sqrt{|\,g|}\left[ i\bar{\Psi}\Gamma^{M}D_{M}\Psi-\eta\bar{\Psi}F(\phi,\chi)\Psi \right]
\end{equation}
\noindent where $\eta$ is the coupling constant between fermions and scalar fields. Here we consider the fields $\phi$ and $\chi$ as background fields. Making the usual change of variable
\begin{equation}\label{cv}
    z=\int^{y}_{0}\mathrm{e}^{-A(y^{\,\prime})}dy^{\,\prime},
\end{equation}
\noindent we get a conformally flat metric
\begin{equation}\label{cm}
    ds^{2}=\mathrm{e}^{2A}\left( \eta_{\mu\nu}dx^{\mu}dx^{\nu} -dz^{2}\right)
\end{equation}
\noindent Using the metric (\ref{cm}) and the representation for gamma matrices $\Gamma^{M}=\left( \mathrm{e}^{-A}\gamma^{\mu},-i\mathrm{e}^{-A}\gamma^{5}\right)$, the equation of motion is
\begin{equation}\label{em}
    \left[ i\gamma^{\mu}\partial_{\mu}+\gamma^{5}(\partial_{z}+2\partial_{z}A)-\eta\mathrm{e}^{A}F(\phi,\chi) \right]\Psi=0.
\end{equation}
\noindent In this stage, we use the general chiral decomposition
\begin{equation}\label{dchiral}
    \Psi(x,z)=\sum_{n}\psi_{L_{n}}(x)\alpha_{L_{n}}(z)+\sum_{n}\psi_{R_{n}}(x)\alpha_{R_{n}}(z),
\end{equation}
\noindent with $\psi_{L_{n}}(x)=-\gamma^{5}\psi_{L_{n}}(x)$ and $\psi_{R_{n}}(x)=\gamma^{5}\psi_{R_{n}}(x)$. With this decomposition $\psi_{L_{n}}(x)$ and $\psi_{R_{n}}(x)$ are the left-handed and right-handed components of the four-dimensional spinor field, respectively. After applying (\ref{dchiral}) in (\ref{em}), and demanding that $i\gamma^{\mu}\partial_{\mu}\psi_{L_{n}}=m_{n}\psi_{R_{n}}$ and $i\gamma^{\mu}\partial_{\mu}\psi_{R_{n}}=m_{n}\psi_{L_{n}}$, we obtain two equations for $\alpha_{L_{n}}$ and $\alpha_{R_{n}}$
\begin{equation}\label{ea1}
    \left[ \partial_{z}+2\partial_{z}A+\eta\mathrm{e}^{A}F(\phi,\chi) \right]\alpha_{L_{n}}=m_{n}\alpha_{R_{n}},
\end{equation}
\begin{equation}\label{ea2}
    \left[ \partial_{z}+2\partial_{z}A-\eta\mathrm{e}^{A}F(\phi,\chi) \right]\alpha_{R_{n}}=-m_{n}\alpha_{L_{n}}.
\end{equation}
\noindent which can be reduced to the Schr\"{o}dinger-like equations for $m_{n}\neq0$.
Inserting the general chiral decomposition (\ref{dchiral}) into the action (\ref{ad}), using (\ref{ea1}) and (\ref{ea2}) and also requiring that the result take the form of the standard four-dimensional action for the massive chiral fermions
\begin{equation}\label{ad2}
    S=\sum_{n}\int d^{4}x\, \bar{\psi}_{n}\left( \gamma^{\mu}\partial_{\mu}-m_{n} \right)\psi_{n},
\end{equation}
\noindent where $\psi_{n}=\psi_{L_{n}}+\psi_{R_{n}}$ and $m_{n}\ge0$, the functions $\alpha_{L_{n}}$ and $\alpha_{R_{n}}$ must obey the following orthonormality conditions
\begin{equation}\label{orto}
    \int_{-\infty}^{\infty}dz\,\mathrm{e}^{4A}\alpha_{Lm}\alpha_{Rn}=\delta_{LR}\delta_{mn}.
\end{equation}

\noindent Implementing the change of variables $\alpha_{L_{n}}=\mathrm{e}^{-2A}L_{n}$ and $\alpha_{R_{n}}=\mathrm{e}^{-2A}R_{n}$, we get
\begin{equation}\label{sleft}
    -L_{n}^{\prime\prime}(z)+V_{L}(z)L_{n}=m_{n}^{2}L_{n},
\end{equation}
\begin{equation}\label{sright}
    -R_{n}^{\prime\prime}(z)+V_{L}(z)R_{n}=m_{n}^{2}R_{n}.
\end{equation}
\noindent where
\begin{eqnarray}
  V_{L}(z) &=& \eta^{2}\mathrm{e}^{2A}F^{2}(\phi,\chi)-\eta\partial_{z}\left( \mathrm{e}^{A}F(\phi,\chi) \right),\label{vefa} \\
  V_{R}(z) &=& \eta^{2}\mathrm{e}^{2A}F^{2}(\phi,\chi)+\eta\partial_{z}\left( \mathrm{e}^{A}F(\phi,\chi) \right)\label{vefb}.
\end{eqnarray}
\noindent Using the expressions $\partial_{z}A=\mathrm{e}^{A(y)}\partial_{y}A$ and $\partial_{z}F=\mathrm{e}^{A(y)}\partial_{y}F$, we can recast the potentials (\ref{vefa}) and (\ref{vefb}) as a function of $y$ \cite{yu}
\begin{eqnarray}
  V_{L}(z(y)) &=& \eta\mathrm{e}^{2A}\left[ \eta F^{2}-\partial_{y}F-F\partial_{y}A(y) \right]\label{vya} \\
  V_{R}(z(y)) &=& V_{L}(z(y))|_{\eta\rightarrow-\eta}\label{vyb}
\end{eqnarray}

\noindent It is worthwhile to note that we can construct the Schr\"{o}dinger potentials $V_{L}$ and $V_{R}$ from (\ref{vya}) and (\ref{vyb}).

Now we focus attention on the motivation of this paper, the calculation of the zero mode. Substituting $m_{n}=0$ in (\ref{ea1}) and (\ref{ea2}) and using $\alpha_{L_{n}}=\mathrm{e}^{-2A}L_{n}$ and $\alpha_{R_{n}}=\mathrm{e}^{-2A}R_{n}$, respectively, we get
\begin{equation}\label{mzL}
    L_{0}\propto \exp \left[-\eta\int_{0}^{z}dz^{\prime}\mathrm{e}^{A(z^{\prime})}F(\phi,\chi) \right]
\end{equation}
\begin{equation}\label{mzR}
    R_{0}\propto \exp \left[\eta\int_{0}^{z}dz^{\prime}\mathrm{e}^{A(z^{\prime})}F(\phi,\chi) \right]
\end{equation}
\noindent This fact is the same to the case of two-dimensional Dirac equation, in this context are called isolated solutions \cite{luis}. At this point is worthwhile to mention that the normalization of the zero mode and the existence of a minimum of the effective potential at the localization on the brane are essential conditions for the problem of fermion localization on the brane.

In order to check the normalization condition (\ref{orto}) for the left-handed fermion zero mode (\ref{mzL}), the integral can be convergent, \textit{i.e}
\begin{equation}\label{cono1}
    \int^{\infty}_{-\infty}dz\exp\left[ -2\eta\int^{z}_{0}dz\,^{\prime}\mathrm{e}^{A(z^{\prime})}F(\phi(z\,^{\prime}),\chi(z\,^{\prime})) \right]<\infty,
\end{equation}
\noindent  and using the expression $dz=\mathrm{e}^{-A(y)}dy$  the integral (\ref{cono1}) can be written as function of $y$ and it becomes
\begin{equation}\label{cono}
    \int^{\infty}_{-\infty}dy\exp\left[ -A(y)-2\eta\int^{y}_{0}dy\,^{\prime}F(\phi(y\,^{\prime}),\chi(y\,^{\prime})) \right]<\infty.
\end{equation}
\noindent This result clearly shows that the normalization of the zero mode is decided by the asymptotic behavior of $F(\phi(y),\chi(y))$. This fact has already been addressed in Ref. \cite{mel} and Ref. \cite{yu}, for various different BPS branes and asymmetric Bloch Branes, respectively.

From (\ref{vya}) and (\ref{vyb}) can be observed that the effective potential profile depends on the choice of $F(\phi(y),\chi(y))$. This fact implies that the existence of a minimum of the effective potential $V_{L}(z(y))$ or $V_{R}(z(y))$ at the localization on the brane is decided by $F(\phi(y),\chi(y))$. This point will be more clear when it is considered a specific Yukawa coupling. Having set up the two essential conditions for the problem of fermion localization on the brane, we are now in a position to choice some specific forms for Yukawa couplings.

\subsection{Case 1: $F(\phi,\chi)=\phi\chi$}

\begin{figure}[ht]
\begin{center}
\includegraphics[width=8cm, angle=0]{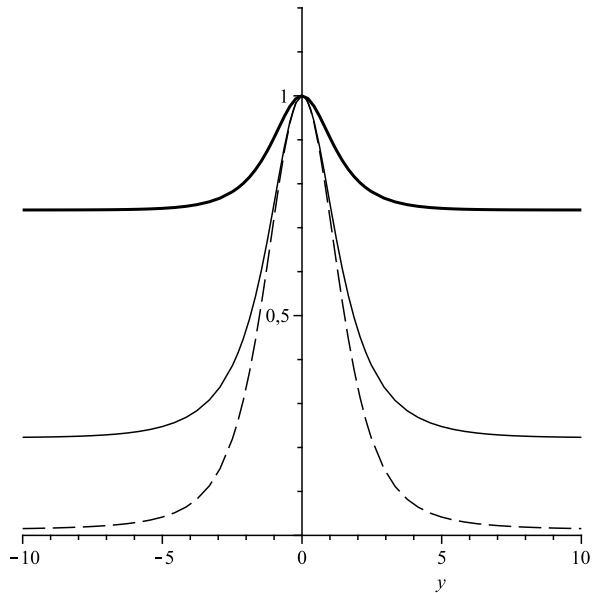}
\end{center}
\par
\vspace*{-0.1cm} \caption{The zero mode of left-handed fermion for the case $F(\phi,\chi)=\phi\chi$, with $\eta=1$, $a=0.20$ (dashed line), $a=0.33$ (thin line) and $a=0.48$ (thick line).} \label{modozero}
\end{figure}

\begin{figure}[ht] 
       \begin{minipage}[b]{0.40 \linewidth}
           \fbox{\includegraphics[width=\linewidth]{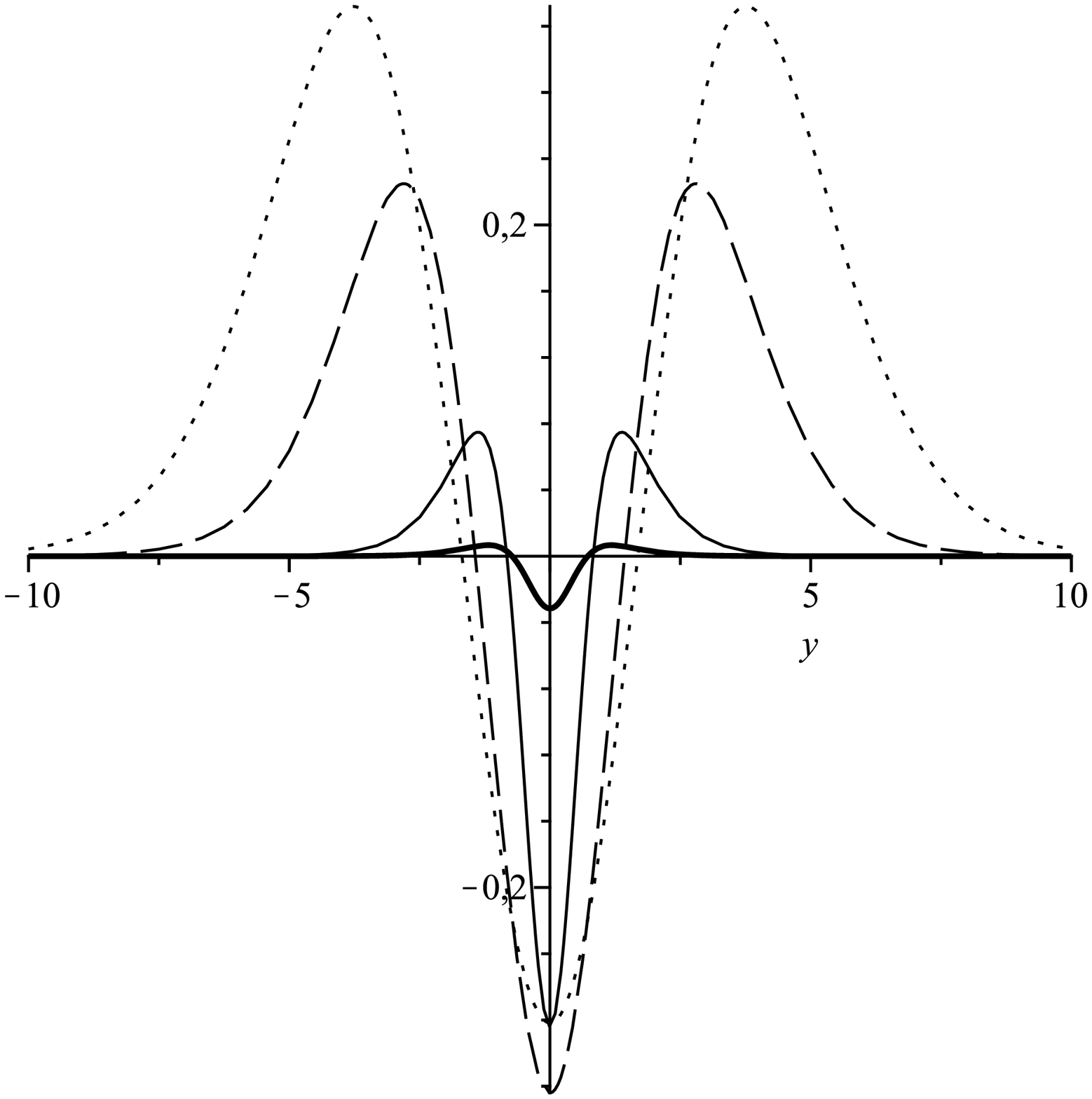}}\\
          \end{minipage}\hfill
       \begin{minipage}[b]{0.40 \linewidth}
           \fbox{\includegraphics[width=\linewidth]{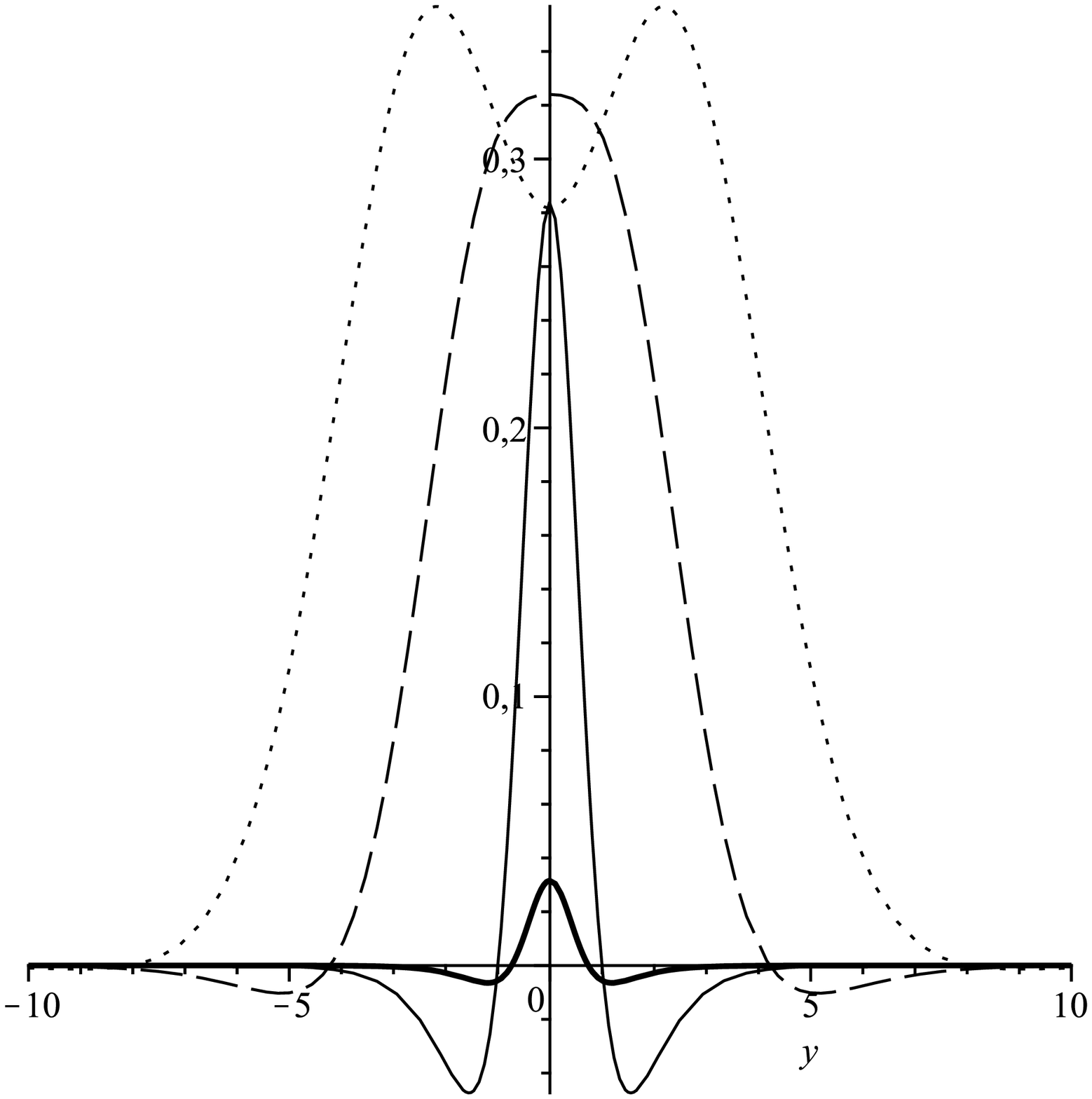}}\\
           \end{minipage}
       \caption{Potential profiles: (a) $V_{L}(y)$ (left) and (b) $V_{R}(y)$ (right) for $\eta=0.5$, $a=0.09$ (dot line), $a=0.15$ (dashed line), $a=0.39$ (thin line) and $a=0.49$ (thick line).}\label{produto}
   \end{figure}

\begin{figure}[ht]
\begin{center}
\includegraphics[width=8cm, angle=0]{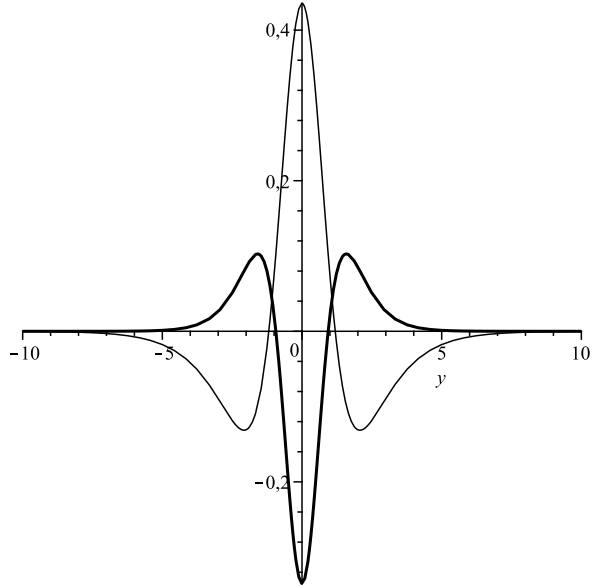}
\end{center}
\par
\vspace*{-0.1cm} \caption{The profiles of the matter energy density (thin line) and $V(y)$ (thick line) for $\eta=0.5$ and $a=1/3$.} \label{devsvlp}
\end{figure}

\begin{figure}[ht]
\begin{center}
\includegraphics[width=8cm, angle=0]{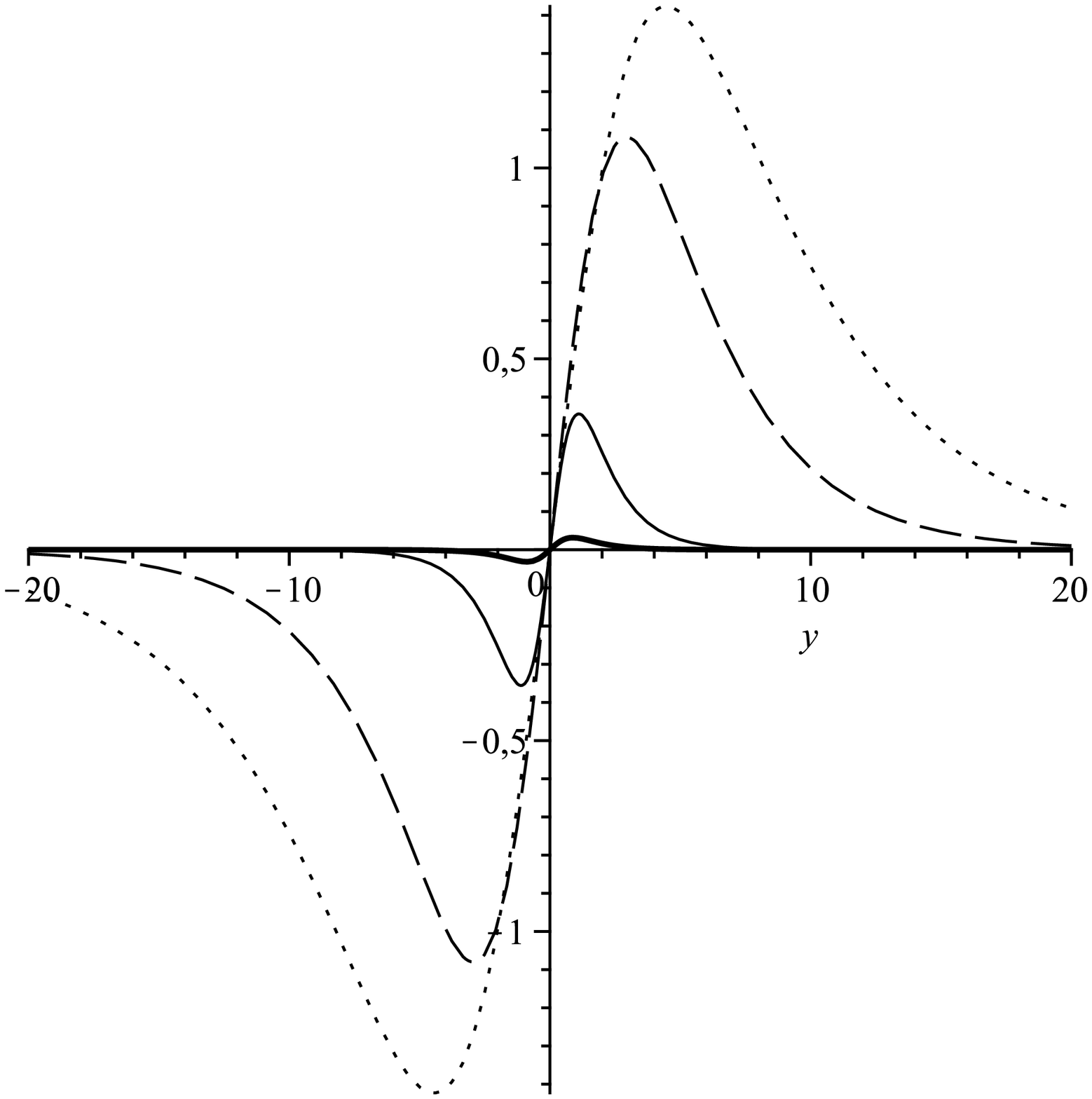}
\end{center}
\par
\vspace*{-0.1cm} \caption{$F(\phi,\chi)=\phi\chi$ as a function of $y$ with $\eta=0.5$, $a=0.09$ (dot line), $a=0.15$ (dashed line), $a=0.39$ (thin line) and $a=0.49$ (thick line).} \label{fproduto}
\end{figure}

This Yukawa coupling has been analyzed in Ref. \cite{casa} for $a=1/3$. In this case, from Eqs. (\ref{sol1}), (\ref{sol2}) and (\ref{sol3}) the integrand in (\ref{cono}) can be expressed as
\begin{equation}\label{int1}
    I=\exp\left\{ -\frac{(1-3a)}{9a}\tanh^{2}(2ay) +\frac{2}{9a}\ln\left[ \cosh (2ay) \right]-\frac{\eta}{a}\sqrt{\frac{1-2a}{a}}\left[ 1-\mathrm{sech}(2ay)  \right] \right\}
\end{equation}
\noindent We follow the same procedure of Ref. \cite{yu}. As $y\rightarrow\infty$,
\begin{equation}
    I\rightarrow\exp\left( \frac{4}{9}\,y \right)\rightarrow\infty
\end{equation}
\noindent and as $y\rightarrow-\infty$,
\begin{equation}
    I\rightarrow\exp\left( -\frac{4}{9}\,y \right)\rightarrow\infty
\end{equation}
\noindent which leads to a non normalizable zero mode. This fact implies that the zero mode of the left-handed fermions can not be localized on the brane, in opposition what was adverted in Ref. \cite{casa}. It is instructive to note that the asymptotic behavior of the integrand is independent of $a$. On the other side, changing $\eta$ by $-\eta$ ($L_{0}\rightarrow R_{0}$) we obtain that the right-handed fermions can not be localized on the brane too. The Fig. \ref{modozero} clearly shows that the zero mode of left-handed fermions is not normalizable. For this case the behavior of the $V_{L}$ and $V_{R}$ potentials are shown in the Fig. \ref{produto} for some values of $a$. Figure \ref{produto}(a) shows that the potential of left-handed fermions, $V_{L}$, is indeed a volcano-like potential. The shapes of the energy density and $V_{L}(y)$ potential for this case are shown in the Fig. \ref{devsvlp} \,for $a=1/3$, as used in Ref. \cite{casa}. The Fig. \ref{devsvlp} clearly shows that the effective potential $V_{L}(y)$ has a minimum at the localization on the brane, but this fact does not guarantee the existence of a normalized zero mode. The behavior of $F(\phi,\chi)=\phi\chi$ as a function of $y$ for some values of $a$ is shown in Fig. \ref{fproduto}. From figure \ref{fproduto} one can see that $F(\phi,\chi)\rightarrow0$ as $y\rightarrow\pm\infty$.

\subsection{Case 2: $F(\phi,\chi)_{\pm}=\phi\pm\chi$}

\begin{figure}[ht] 
       \begin{minipage}[b]{0.40 \linewidth}
           \fbox{\includegraphics[width=\linewidth]{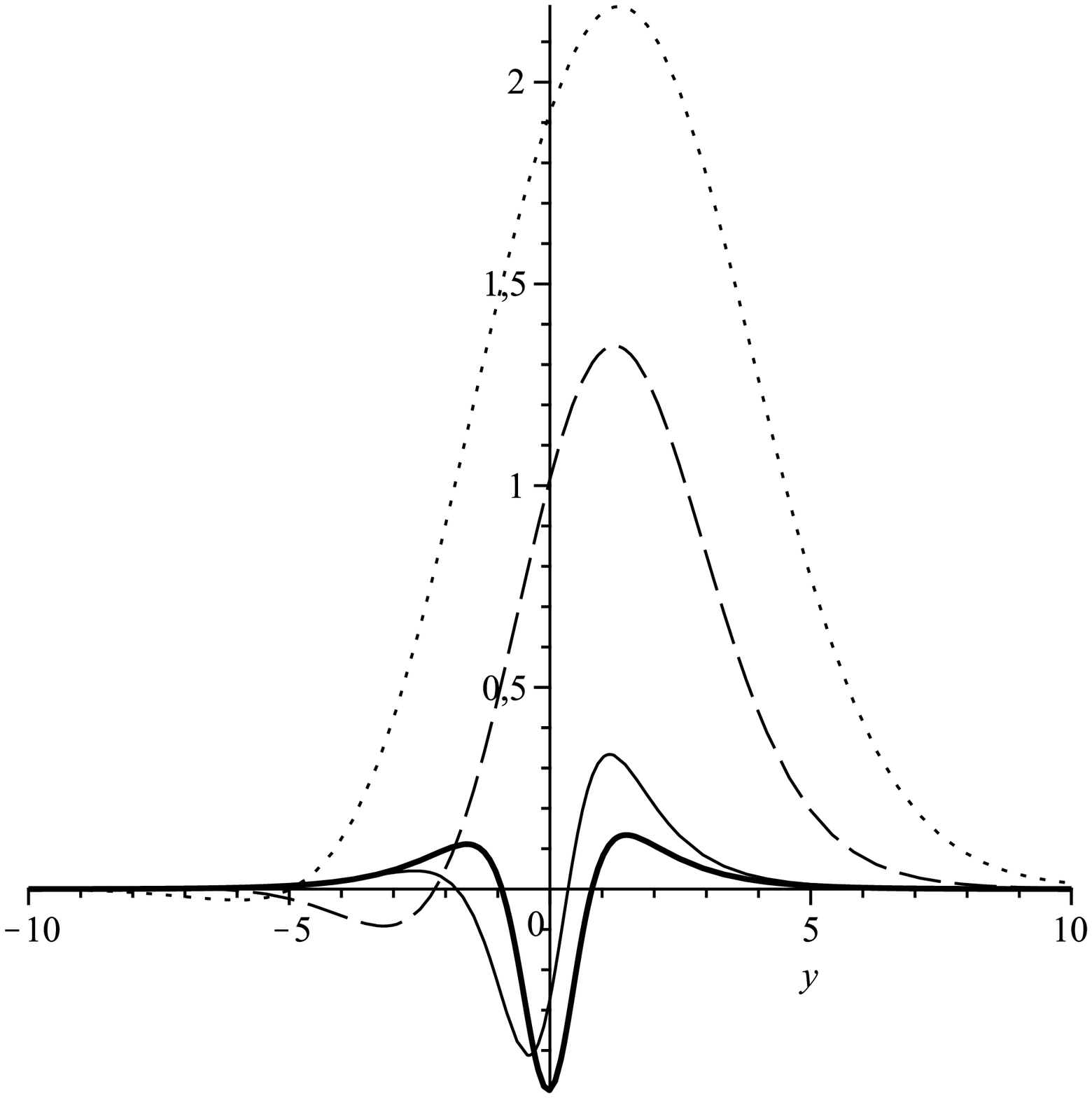}}\\
          \end{minipage}\hfill
       \begin{minipage}[b]{0.40 \linewidth}
           \fbox{\includegraphics[width=\linewidth]{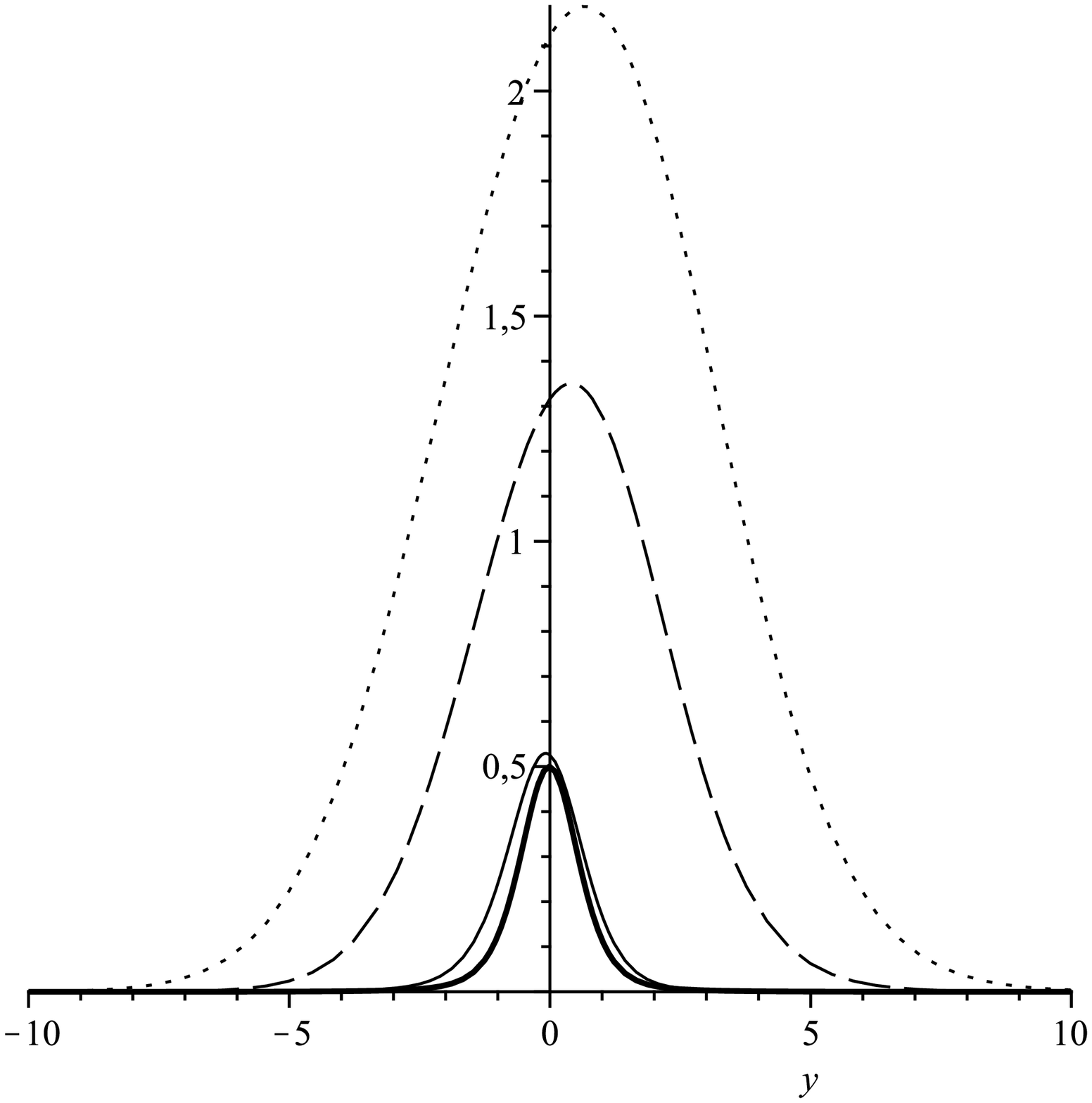}}\\
           \end{minipage}
       \caption{Potential profile: (a) $(V_{L}(y))_{+}$ (left) and (b) $(V_{R}(y))_{+}$ (right) for $\eta=0.5$, $a=0.09$ (dot line), $a=0.15$ (dashed line), $a=0.39$ (thin line) and $a=0.49$ (thick line).}\label{suma}
   \end{figure}

\begin{figure}[ht] 
       \begin{minipage}[b]{0.40 \linewidth}
           \fbox{\includegraphics[width=\linewidth]{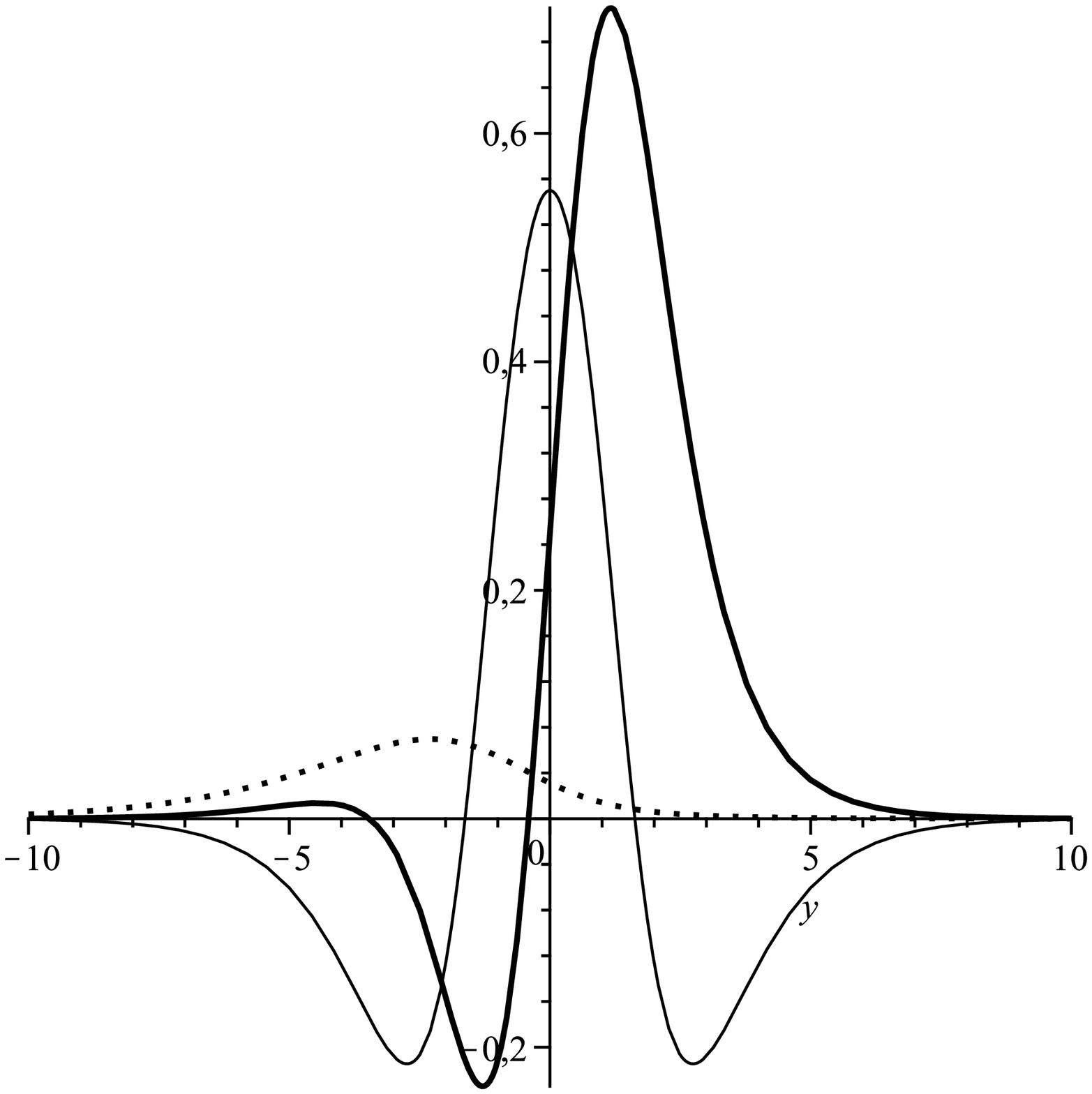}}\\
          \end{minipage}\hfill
       \begin{minipage}[b]{0.40 \linewidth}
           \fbox{\includegraphics[width=\linewidth]{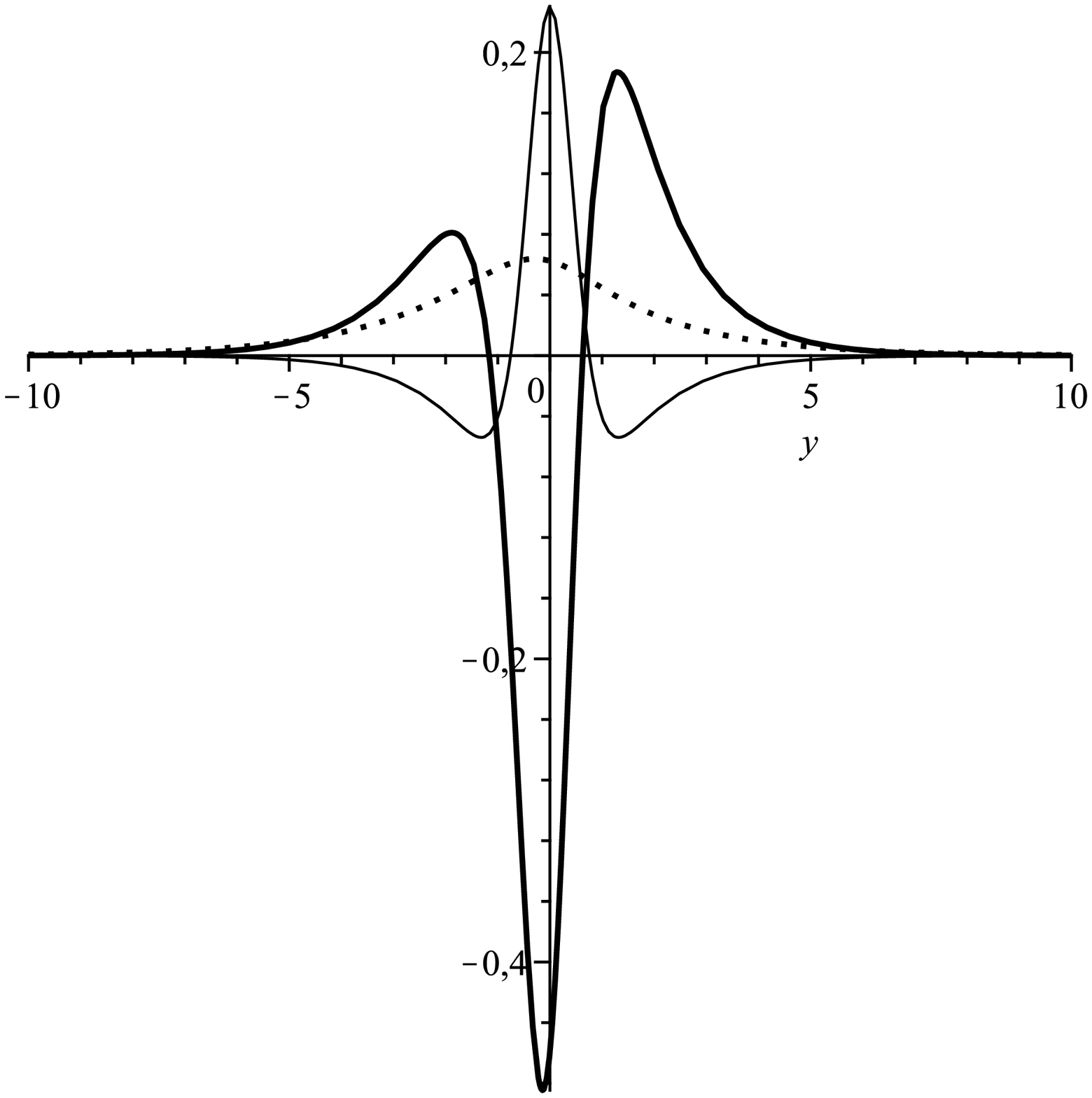}}\\
           \end{minipage}
       \caption{The profiles of the energy density (thin line), $(V_{L}(y))_{+}$ (thick line) and zero mode (dot line) for $\eta=0.5$; (a) $a=0.25$ (left) and (b) $a=0.48$ (right).}\label{devsvls}
   \end{figure}

For this case the integrand in (\ref{cono}) can be expressed as
\begin{equation}\label{int2}
    I=\exp\left\{ -\frac{1}{a}\left(\eta-\frac{2}{9}\right)\ln\left[\cosh(2ay) \right]-\frac{(1-3a)}{9a}\tanh^{2}(2ay) \mp\frac{\eta}{a}\sqrt{\frac{1-2a}{a}}\arctan\left[\sinh(2ay) \right] \right\}.
\end{equation}
\noindent The behavior of (\ref{int2}) as $y\rightarrow\infty$ is given by
\begin{equation}
    I\rightarrow\exp\left[ -2\left(\eta-\frac{2}{9} \right)\,y \right]\rightarrow0,\qquad \mathrm{for} \quad \eta>2/9
\end{equation}
\noindent and as $y\rightarrow-\infty$, is
\begin{equation}
    I\rightarrow\exp\left[ 2\left(\eta-\frac{2}{9} \right)\,y \right]\rightarrow0,\qquad \mathrm{for} \quad \eta>2/9
\end{equation}
\noindent This result clearly shows that the zero mode of the left-handed fermions is normalized only for $\eta>2/9$. Now, under the change $\eta\rightarrow-\eta$ ($L_{0}\rightarrow R_{0}$) we obtain that the right-handed fermions can not be a normalizable zero mode. The shape of the potentials for $F(\phi,\chi)_{+}$ are shown in Fig. \ref{suma} for some values of $a$. Figure \ref{suma}(a) shows a well structure that grows and tend to around $y=0$ as $a\rightarrow1/2$. Figure \ref{suma}(b) shows that the potential $(V_{R}(y))_{+}$ is always positive, therefore, the potential can not trap any bound fermions with right chirality. This kind of potentials has not appeared in previous studies. The shapes of the energy density, $(V_{L}(y))_{+}$ potential and zero mode for this case are shown in Fig. \ref{devsvls}. The Fig. \ref{devsvls}(a) ($a=0.25$) shows that the effective potential $(V_{L}(y))_{+}$ has a minimum outside the localization of the brane, as a consequence the normalizable zero mode is not localized on the brane. On the other hand, the Fig. \ref{devsvls}(b) ($a=0.48$) clearly shows that the effective potential $(V_{L}(y))_{+}$ has a minimum at the localization of the brane. Therefore, this result clearly shows that the zero mode of the left-handed fermions is localized on the brane only as $a\rightarrow1/2$.

\begin{figure}[ht]
\begin{center}
\includegraphics[width=8cm, angle=0]{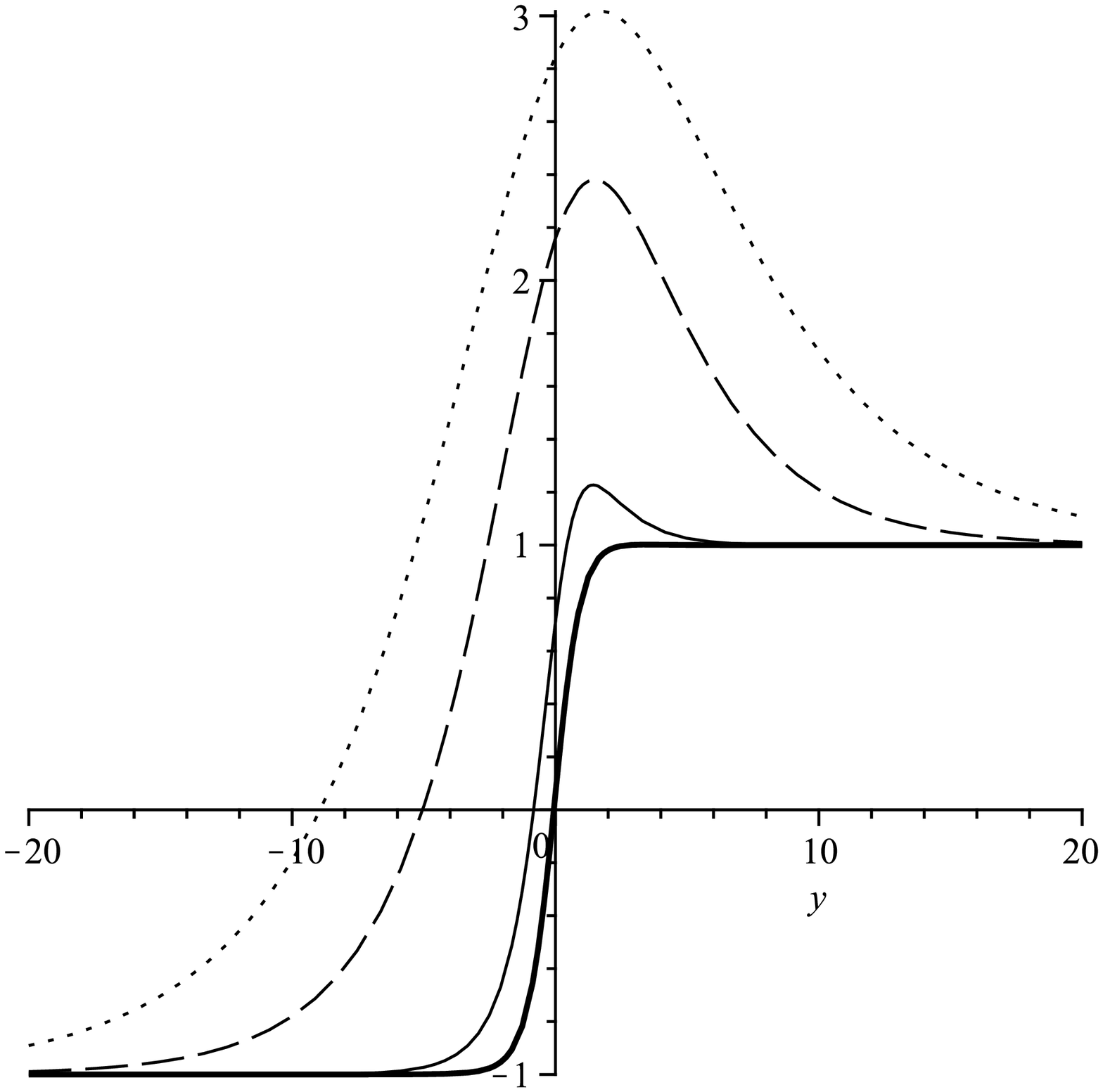}
\end{center}
\par
\vspace*{-0.1cm} \caption{$F(\phi,\chi)_{+}=\phi+\chi$ as a function of $y$ with $\eta=0.5$, $a=0.09$ (dot line), $a=0.15$ (dashed line), $a=0.39$ (thin line) and $a=0.49$ (thick line).} \label{fsuma}
\end{figure}

The behavior of the potentials for $F(\phi,\chi)_{-}$ can be written out easily by replacing $(V_{L})_{-}=(V_{L}(-y))_{+}$ and $(V_{R})_{-}=(V_{R}(-y))_{+}$. The behavior of $F(\phi(y),\chi(y))_{+}$ is shown in Fig. \ref{fsuma}.

\subsection{Case 3: $F(\phi,\chi)=\chi-\phi$}

For this case the integrand in (\ref{cono}) can be expressed as
\begin{equation}\label{int3}
    I=\exp\left\{\frac{1}{a}\left(\eta+\frac{2}{9}\right)\ln\left[\cosh(2ay) \right]-\frac{(1-3a)}{9a}\tanh^{2}(2ay) -\frac{\eta}{a}\sqrt{\frac{1-2a}{a}}\arctan\left[\sinh(2ay) \right] \right\}.
\end{equation}
\noindent The behavior of (\ref{int3}) as $y\rightarrow\infty$ is
\begin{equation}
    I\rightarrow\exp\left[ 2\left(\eta+\frac{2}{9} \right)\,y \right]\rightarrow\infty
\end{equation}
\noindent and as $y\rightarrow-\infty$, is given by
\begin{equation}
    I\rightarrow\exp\left[-2\left(\eta+\frac{2}{9} \right)\,y \right]\rightarrow\infty
\end{equation}
\noindent which leads to a non normalizable zero mode. The zero mode of the left-handed fermions can not be localized on the brane. Otherwise, the change $\eta\rightarrow-\eta$ ($L_{0}\rightarrow R_{0}$) allowed us to conclude that the right-handed fermions can be localized on the brane on the condition that $\eta>2/9$ and $a\rightarrow1/2$. The behavior of the potentials for this $F(\phi,\chi)$ can be written out easily by replacing $V_{L}=(V_{R}(-y))_{+}$ and $V_{R}=(V_{L}(-y))_{+}$.

\section{Conclusions}

We have reinvestigated the localization problem of fermions on two-field thick branes (Bloch brane model). We showed that the simplest Yukawa coupling $F(\phi,\chi)=\phi\chi$ does not support the localization of fermions on the brane, as incompletely argued in Ref. \cite{casa}. This fact is a consequence of the absence of a normalized zero mode. We showed that the zero mode of left-handed fermions for the Yukawa coupling $F(\phi,\chi)_{\pm}=\phi\pm\chi$ is normalizable under the condition $\eta >2/9$ and it can be trapped on the brane only for $a\rightarrow1/2$, because the effective potential have a minimum at the localization of the brane. On the other hand, the zero mode of right-handed is not normalizable, this result can also be validated from the behavior of the potential $(V_{R})_{\pm}$ (Figure \ref{suma}(b)). In the same way, we also showed that the zero mode right-handed fermions can be localized on the brane for $\eta >2/9$ and $a\rightarrow1/2$ for $F(\phi,\chi)=\chi-\phi$. At this point, it is worthwhile to mention that the limit $a\rightarrow1/2$ changes the two-field solution to the one-field solution \cite{ba}. In this context, we believe that an interesting line of investigation could follow \cite{aug}, which construct effective models with only one scalar field from models with two interacting scalar fields.

We can conclude that the normalization of the zero mode and the existence of a minimum of the effective potential at the localization on the brane are essential conditions for the problem of fermion localization on the brane. The normalization of the zero mode is decided by the asymptotic behavior of $F(\phi,\chi)$ and also the presence of $F(\phi,\chi)$ is an essential ingredient for the effective potential profile. Therefore, the behavior of $F(\phi,\chi)$ plays a leading role for the fermion localization on the brane. This work completes the analyze of the research in Ref. \cite {casa}, because in that work does not analyze the zero mode in full detail. An interesting issue concerns the natural extension of the present work to the case of massive modes using the other kinds of Yukawa couplings which support normalizable zero mode presented in this paper in order to investigate possible effects on the resonances modes and bear out the main conclusions of Ref. \cite{casa}.

\begin{acknowledgments}
The author would like to thank Professor Dr. Antonio S. de Castro and Professor Marcelo B. Hott. Thanks also go to the referee for useful comments and suggestions. This work was supported by means of funds provided by CAPES.
\end{acknowledgments}


\begin{thebibliography}{99}

\bibitem{casa} C.A.S. Almeida, R. Casana, M.M. Ferreira Jr. and A.R. Gomes, Phys. Rev. D \textbf{79}, 125022 (2009).

\bibitem{ba} D. Bazeia and A.R. Gomes, J. High Energy Phys. \textbf{05} 012 (2004).

\bibitem{du} A. de Souza Dutra, A.C. Amaro de Faria and M. Hott, Phys. Rev. D \textbf{78}, 043526 (2008).

\bibitem{mel} A. Melfo, N. Pantoja and J.D. Tempo, Phys. Rev. D \textbf{73}, 044033 (2006).

\bibitem{yu} Z.-H. Zhao, Y.-X. Liu and H.-T Li, Class. Quantum Grav. \textbf{27}, 185001 (2010).

\bibitem{liu} Y.-X. Liu, C.-E. Fu, L. Zhao, Y.-S. Duan, Phys. Rev. D \textbf{80}, 065020 (2009).
\bibitem{hat} M. Hatsuda and M. Sakaguchi, Nucl. Phys. B \textbf{577}, 183 (2000).

\bibitem{ran} S. Randjbar-Daemi and M. Shaposhnikov, Phys. Lett. B \textbf{492}, 361 (2000).

\bibitem{mou} S. Mouslopoulos, J. High Energy Phys. \textbf{05}, 038 (2001).

\bibitem{dub} S.L. Dubovsky, V.A. Rubakov and P.G. Tinyakov, Phys. Rev. D \textbf{62}, 105011 (2000).

\bibitem{rin} C. Ringeval, P. Peter and J.-P. Uzan, Phys. Rev. D \textbf{65}, 044016 (2002).

\bibitem{bie} W. Bietenholz, A. Gfeller and U.-J. Wiese, J. High Energy Phys. \textbf{10}, 018 (2003).

\bibitem{kol} R. Koley and S. Kar, Classical Quantum Gravity \textbf{22}, 753 (2005).

\bibitem{gib} G. Gibbons, K. Maeda and Y. Takamizu, Phys. Lett. B \textbf{647}, 1 (2007).

\bibitem{xia} Y.-X. Liu, L.-D. Zhang, L.-J. Zhang and Y-.S. Duan, Phys. Rev. D \textbf{78}, 065025 (2008).

\bibitem{xia2} Y.-X. Liu, X.-H. Zhang, L.-D. Zhang and Y-.S. Duan, J. High Energy Phys. \textbf{02}, 067 (2008).

\bibitem{xia3} Y.-X. Liu, L.-D. Zhang, S.-W. Wei and Y-.S. Duan, J. High Energy Phys. \textbf{08}, 041 (2008).

\bibitem{dav} R. Davies and D.P. George, Phys. Rev. D \textbf{76}, 104010 (2007).

\bibitem{wol} O. DeWolfe, D.Z. Freedman, S.S. Gubser and A. Karch, Phys. Rev. D \textbf{62}, 046008 (2000).

\bibitem{ba2} D. Bazeia, M.J. dos Santos and R.F. Ribeiro, Phys. Lett. A \textbf{208}, 84 (1995).
\bibitem{luis} A.S. de Castro and M.B. Hott, Phys. Lett. A \textbf{351}, 379 (2006); L.B. Castro and A.S. de Castro, J. Phys. A: Math. theor. \textbf{40}, 263 (2007); L.B. Castro and A.S. de Castro, Phys. Scr. \textbf{75}, 170 (2007); L.B. Castro and A.S. de Castro, Int. J. Mod. Phys. E \textbf{16}, 2998 (2007); L.B. Castro, A.S. de Castro and M.B. Hott, Int. J. Mod. Phys. E \textbf{16}, 3002 (2007); L.B. Castro and A.S. de Castro, Phys. Scr. \textbf{77}, 045007 (2008).

\bibitem{aug} A.E.R. Chumbes and M.B. Hott, Phys. Rev. D \textbf{81}, 045008 (2010).

\end{thebibliography}
\end{document}